\documentclass[a4paper,11pt]{article}
\usepackage{graphicx}
\usepackage{subfig}
\usepackage{color}
\usepackage{amsfonts,amssymb,latexsym,amsmath}
\usepackage[section]{placeins}
\usepackage{a4wide}
\usepackage{bm}
\usepackage{array} 
\usepackage{epstopdf}
\usepackage{cite}
\usepackage{float}
\usepackage{hyperref}
\usepackage{array}
\usepackage{adjustbox}
\usepackage{colortbl}
\usepackage{tabularx}
\usepackage{xcolor}

\title{Spin-averaged $B_c$ Spectrum in a Cornell-type Potential Using VMC Baseline and GFMC Evolution}
\author{Tarik Akan\thanks{tarik.akan@bozok.edu.tr} \\
Physics Department, Yozgat Bozok University, 66100 Yozgat, Turkey}
\date{\today}

\begin{document}

\maketitle
\begin{abstract}
In this work, the spin-averaged $B_c$ spectrum is computed in a naive Cornell framework, treating the meson as a nonrelativistic system in a spin-independent potential. The Cornell parameters are calibrated directly to the spin-averaged $B_c$ tower by anchoring the $1S$ centroid and scanning a grid in $(\sigma,\kappa)$, with the additive constant $V_0$ fixed at each point by the experimental ground state mass. The spectrum is obtained with a two stage Monte Carlo approach. Variational Monte Carlo (VMC) provides optimized radial trial states with the desired nodal pattern. Fixed node Green's function Monte Carlo (GFMC) then projects the corresponding ground state energies for each $(n,\ell)$ channel. Controlled scans over the GFMC time step, projection time, walker population, and radial grid identify plateau regions where discretization and projection systematics are quantitatively under control. At a representative best point in the low-RMSE valley, the predicted spin-averaged masses agree with the experimental centroids at the level of a few tens of MeV, and the fitted Cornell parameters are consistent with canonical heavy quarkonium analyses.

\end{abstract}

\section{Introduction}

Heavy quark bound states offer a comparatively clean environment in which to test ideas about nonrelativistic QCD and to study how the dynamics interpolates between the perturbative and confining regimes. Beyond conventional quarkonia, modern nonrelativistic treatments now encompass baryons, multihadron systems, and even composite dark matter candidates, providing a broad arena in which heavy quark dynamics can be constrained and cross checked~\cite{godfrey1985mesons,godfrey2004spectroscopy,ebert2011spectroscopy,akan2025predicting,mutuk2019cornell,assi2023baryons,soni2018qq}. Within this family, the $B_c$ meson is particularly informative. Because it is composed of two heavy quarks with different flavors, its spectrum is governed by a mixed reduced mass and is simultaneously sensitive to the short distance Coulombic interaction and the long distance confining rise of the potential. The low-lying spin-averaged levels therefore provide a compact set of benchmarks against which any proposed description of heavy quark dynamics can be confronted.

Experimentally, the $B_c$ ground state was first observed in hadron collider data~\cite{cdf1998observation,abazov2008measurement} and its basic properties, such as mass and lifetime, have since been measured with increasing precision~\cite{lhcb2015measurement,aaij2018measurement,aaij2023study,beringer2012review,patrignani2016review}. On the theory side, the $B_c$ spectrum has been studied in a variety of frameworks, including relativistic quark models~\cite{ebert2003properties,godfrey2004spectroscopy,ebert2011spectroscopy,devlani2014masses,monteiro2017cb,chaturvedi2022b}, lattice QCD calculations of the ground state~\cite{allison2005mass}, Dyson-Schwinger and Bethe-Salpeter approaches~\cite{Chen_2020}, and potential model analyses of higher excitations~\cite{li2023higher,li2023spectroscopic}. Taken together, these works provide a detailed but method-dependent picture of the $B_c$ tower and highlight the usefulness of this system as a bridge between charmonium and bottomonium.

Among potential model descriptions, the Cornell potential has played a central role for several decades. In its original formulation it encodes the static interaction between a heavy quark and antiquark as the sum of a $1/r$ term, motivated by one gluon exchange, and a linear term that reproduces the long-distance behavior seen in lattice simulations and flux tube pictures of confinement~\cite{eichten1978charmonium,eichten1980charmonium,godfrey1985mesons}. Phenomenologically, this potential succeeds in organizing a wide range of spectroscopy, transition, and decay data across charmonium, bottomonium, and heavy-light mesons~\cite{eichten1994mesons,godfrey2004spectroscopy,ebert2003properties}. More recent studies have mapped out the allowed parameter space of Cornell-type potentials directly from QCD motivated considerations~\cite{solomko2023cornell,pathak2022parameterisation,sreelakshmi2022mass,ahmad2025charmonium}, reinforcing the view that a suitably calibrated Cornell Hamiltonian can serve as a minimal, QCD compatible baseline for heavy quark spectroscopy.

This close connection to QCD motivates a fresh look at the $B_c$ system within the same naive Cornell framework. In the present work, the $B_c$ meson is treated as a nonrelativistic two-body problem with a central, spin-independent Cornell potential supplemented by an additive constant that fixes the overall energy origin. This analysis focuses on the spin-averaged $B_c$ tower ($1S$, $1P$, $1D$, $2S$, $2P$, $3S$, $3P$, $4S$), using experimental centroids where available and enlarging the associated uncertainties when multiplets are incomplete~\cite{beringer2012review,patrignani2016review}. The aim is deliberately narrow: the anaysis asks how far a minimal Cornell Hamiltonian, with parameters calibrated directly to the $B_c$ spectrum, can account for the observed pattern of radial and orbital excitations and seek to identify a region in $(\sigma,\kappa,V_0)$ space that is consistent both with the data and with canonical heavy quarkonium fits~\cite{eichten1994mesons,soni2018qq,li2023higher,li2023spectroscopic}.

The main aim of the present work is not to propose a new static potential for the $B_c$ system, but to establish a numerically controlled Monte Carlo framework for the spin averaged $B_c$ spectrum within a minimal Cornell type Hamiltonian. The spectrum of the Cornell Hamiltonian is obtained with a combination of variational Monte Carlo (VMC) and Green’s function Monte Carlo (GFMC) in imaginary time~\cite{assi2023baryons, assi2024tetraquarkssufficientlyunequalmassheavy,sorella1998green}. VMC provides compact radial trial states with the correct nodal structure and yields reference energies that are free of time step artifacts, while fixed node GFMC evolution refines these energies by projecting onto the lowest state compatible with each nodal pattern. In this way, the calculation complements existing $B_c$ studies based on semi-analytic solution methods such as the asymptotic iteration method~\cite{ciftci2003asymptotic,kumar2013asymptotic} and on modern neural network solvers for Cornell type potentials~\cite{mutuk2019cornell,akan2025predicting}. Particular attention is paid to how the Monte Carlo control parameters influence the calibration of the Cornell potential, since the same numerical machinery can later be reused for spin-dependent and relativistic corrections.

The potential parameters are extracted by scanning a grid in $(\sigma,\kappa)$, fixing $V_0$ at each point by anchoring the $1S$ centroid to experiment, and comparing the resulting spin-averaged masses with the experimental centroids across the $B_c$ tower. A single root mean square error (RMSE) built from absolute masses identifies an anticorrelated valley in parameter space where the spectrum is well reproduced. Within this valley, candidate parameter sets are further discriminated by inspecting residuals and requiring that the entire $B_c$ tower be described without systematic drifts with excitation. The resulting ``best valley'' parameters can then be compared with those inferred in other heavy quark systems~\cite{eichten1994mesons,solomko2023cornell,pathak2022parameterisation}, providing an additional consistency check of the naive Cornell picture in a mixed-flavor heavy meson.

Therefore, the paper should be read as a computationally controlled spectroscopy baseline for the $B_c$ system, with emphasis on method validation and parameter stability. Within this scope in mind, the rest of the paper is organized as follows. Section~\ref{sec:hamiltonian} reviews the nonrelativistic Schrödinger problem for the $B_c$ meson, introduces the Cornell potential, and describes the structure of the radial trial states used in VMC. Section~\ref{sec:methodology} summarizes the VMC and GFMC algorithms, the treatment of time step and projection systematics, and the numerical diagnostics used to define the plateau regions. Section~\ref{sec:calibration} presents the calibration of $(\sigma,\kappa,V_0)$ to the spin-averaged $B_c$ tower, including the RMSE maps and representative spectra. Section~\ref{sec:results} discusses the resulting $B_c$ masses in comparison with other potential models and lattice determinations and with experimental data, and the paper concludes with a brief outlook on extensions to spin-dependent splittings, relativistic corrections, and alternative static potentials.

\section{Schrödinger Equation, Potential, and Trial States}
\label{sec:hamiltonian}

In the nonrelativistic, spin-independent limit ($\hbar=c=1$), the $B_c$ meson reduces to a two body problem with reduced mass $\mu=m_b m_c/(m_b+m_c)$. Writing $\Psi(\mathbf{r})=R_{n\ell}(r)\,Y_{\ell m}(\hat{\mathbf{r}})$ and $u_{n\ell}(r)=r\,R_{n\ell}(r)$, the radial bound states satisfy
\begin{equation}\label{eq:radial}
-\frac{1}{2\mu}\,\frac{d^2 u_{n\ell}}{dr^2}
+\Bigl[\frac{\ell(\ell+1)}{2\mu\,r^2}+V(r)\Bigr]\,u_{n\ell}(r)
=E_{n\ell}\,u_{n\ell}(r),
\end{equation}
with regular behavior $u_{n\ell}(r)\sim r^{\ell+1}$ as $r\to 0$ and negligible amplitude at a large but finite $r_{\max}$ so that $u_{n\ell}(r_{\max})\simeq 0$. Closely related radial formulations and solution strategies for heavy quark-antiquark systems can be found, for example, in Refs.~\cite{ciftci2003asymptotic,kumar2013asymptotic}. Physical masses are assembled as
\begin{equation}\label{eq:mass}
M_{n\ell}=m_b+m_c+E_{n\ell}.
\end{equation}

The interaction is modeled by a central Cornell potential,
\begin{equation}\label{eq:veff}
V(r)=\sigma\,r-\frac{\kappa}{r}+V_0,
\end{equation}
which serves as the working potential in all calculations. Here $\sigma$ governs the long-distance confining rise and largely sets the spacing of radial and orbital excitations, while $\kappa$ encodes the short distance attraction associated with one gluon exchange and pulls $S$ waves more strongly toward the origin than higher $\ell$ states. This Cornell form and its variants have been used extensively in heavy quark spectroscopy since the classic charmonium and $b\bar c$ studies \cite{eichten1978charmonium,eichten1980charmonium,godfrey1985mesons,eichten1994mesons}, and its parameter space has been revisited more recently in QCD motivated analyses~\cite{solomko2023cornell,pathak2022parameterisation,sreelakshmi2022mass,ahmad2025charmonium,li2023higher,li2023spectroscopic}. The centrifugal term in Eq.~\eqref{eq:radial} is kinematic, arising from separation of variables in the kinetic energy, and lifts $P$- and $D$-wave levels relative to $S$ waves by repelling probability from small $r$. The constant $V_0$ fixes the overall energy origin for a given flavor assignment; splittings are independent of $V_0$, whereas absolute masses depend on it. In natural units with $r$ measured in $GeV^{-1}$, the parameters carry dimensions $[\sigma]=GeV^2$, $[\kappa]=1$, $[V_0]=GeV$, and $[\mu]=GeV$. Near the origin the Coulomb and centrifugal pieces control the shape of the wave function (a cusp for $\ell=0$ and power law suppression for $\ell\ge 1$), while at large $r$ the linear term dominates and determines the overall ladder of excitations.

Trial states are chosen as smooth, normalizable states tailored to each $(n,\ell)$ channel rather than fixed closed form bases. The states respect the near origin behavior implied by Eq.~\eqref{eq:radial} and the Cornell potential, incorporate the standard $r^\ell$ factor for $\ell>0$, and decay sufficiently gently that reflections from $r_{\max}$ are negligible on the scales of interest.Excited states are obtained by inserting the required number of simple radial nodes into a common template, so the intended structure is enforced without overparameterization. A small set of interpretable parameters, such as an overall range scale and, when needed, the positions of the radial nodes, is adjusted through a short preliminary variational fit to capture the size and structure of each state while keeping the trial space compact. Similar physics motivated trial or basis functions underlie many quark model and potential model treatments of $B_c$ and heavy quarkonia~\cite{ebert2003properties,godfrey2004spectroscopy,ebert2011spectroscopy,devlani2014masses,soni2018qq,mutuk2019cornell,akan2025predicting}.

Orthogonality within a fixed $(\ell,m)$ channel is enforced by projection. If $|\Psi_0\rangle,\ldots,|\Psi_k\rangle$ denote already obtained orthonormal levels, a boundary compatible seed $|\tilde\Psi_t\rangle$ is mapped to the orthogonal complement according to
\begin{equation}\label{eq:proj}
|\Psi_t\rangle=\Bigl(\mathbb{I}-\sum_{j=0}^{k}|\Psi_j\rangle\langle\Psi_j|\Bigr)\,|\tilde\Psi_t\rangle,
\end{equation}
then normalized and used both for a variational energy estimate and as a stable guide for imaginary-time propagation. This Gram-Schmidt type construction is the standard way to build excited states in variational and projector Monte Carlo treatments and is conceptually consistent with earlier excited state studies based on potential models and semi-analytic methods~\cite{ciftci2003asymptotic,kumar2013asymptotic,devlani2014masses,li2023higher,li2023spectroscopic}. It yields compact trial functions for the $1S$, $1P$, $1D$, $2S$, $2P$, $3S$, $3P$, and $4S$ channels and keeps the connection to boundary behavior explicit while maintaining numerical stability in the subsequent projector evolution.

\section{Methodology}
\label{sec:methodology}

The calculation proceeds in two complementary stages that address the same radial Hamiltonian,
\begin{equation}
H=-\frac{1}{2\mu}\frac{d^2}{dr^2}+\frac{\ell(\ell+1)}{2\mu\,r^2}+V(r), 
\qquad 
V(r)=\sigma r-\frac{\kappa}{r}+V_0,
\end{equation}
with regular behavior at $r=0$ and negligible amplitude at a large but finite $r_{\max}$. Calculations are performed at fixed $(\ell,m)$, and excited states are constructed by enforcing orthogonality to previously obtained levels within the same $(\ell,m)$ subspace. In practice the angular dependence is carried entirely by spherical harmonics and factored out, so that Monte Carlo sampling is performed only in the radial coordinate with measure $r^2dr$. Closely related projector Monte Carlo strategies for nonrelativistic QCD few body systems can be found, for example, in Refs.~\cite{assi2023baryons,assi2024tetraquarkssufficientlyunequalmassheavy}.

The first stage is a variational Monte Carlo (VMC) calculation. For a normalized trial state $\Psi(\mathbf{r};\boldsymbol{\theta})$ in a chosen $(n,\ell)$ sector, the variational energy is
\begin{equation}
E[\Psi]=\frac{\langle \Psi|H|\Psi\rangle}{\langle \Psi|\Psi\rangle}
=\frac{\int d^3\mathbf{r}\,|\Psi(\mathbf{r})|^2\,E_L(\mathbf{r})}{\int d^3\mathbf{r}\,|\Psi(\mathbf{r})|^2},
\qquad
E_L(\mathbf{r})=\frac{H\Psi(\mathbf{r})}{\Psi(\mathbf{r})}.
\end{equation}
Sampling from $|\Psi|^2$ with a standard Metropolis kernel yields unbiased estimates of $E[\Psi]$ once autocorrelations are under control. A good trial state has a comparatively narrow local energy distribution, as is standard practice in VMC applications to nonrelativistic QCD and hadronic systems~\cite{assi2023baryons}.

Parameter updates are guided by derivatives of the energy,
\begin{equation}
\frac{\partial E}{\partial \theta_i}
=
2\Big\langle \big(E_L-E\big)\,O_i \Big\rangle_{|\Psi|^2},
\end{equation}
where
\begin{equation}
O_i(\mathbf{r})=\frac{\partial}{\partial \theta_i}\ln \Psi(\mathbf{r};\boldsymbol{\theta}),
\end{equation}
and only a small, physically interpretable set of parameters is varied to avoid ill conditioning.In practice, an optimal linear combination of the $O_i$ is also subtracted from $E_L$ when forming the right hand side, which leaves the mean gradient unchanged while reducing its variance and making the optimization more stable. Similar ideas underlie correlated sampling and variance reduction strategies in other nonrelativistic QCD Monte Carlo studies~\cite{assi2023baryons,assi2024tetraquarkssufficientlyunequalmassheavy}.

Excited states are obtained by projecting trial seeds onto the orthogonal complement of lower levels in the same $(\ell,m)$ channel and then applying the same optimization. Starting from a simple seed with the desired number of nodes, components along previously optimized states are subtracted, the result is normalized, and the projected state is reoptimized. This projection is reapplied periodically, so that slow loss of orthogonality due to Monte Carlo noise does not accumulate. The outcome of the VMC stage is a collection of $\Delta\tau$ independent reference energies and compact guiding functions with the correct nodal structure and a well behaved local energy distribution, providing the fixed node templates for the subsequent projector evolution, in close analogy to the role of trial states in Refs.~\cite{assi2023baryons,assi2024tetraquarkssufficientlyunequalmassheavy}.

The second stage is an imaginary time projection using Green’s function Monte Carlo (GFMC), which removes residual variational bias within each fixed nodal sector. Writing $\tau=it$, the imaginary time evolution
\begin{equation}
-\frac{\partial \Phi(\mathbf{r},\tau)}{\partial \tau}
=
\big(H-E_T\big)\,\Phi(\mathbf{r},\tau),
\qquad
\Phi(\mathbf{r},\tau)=e^{-\tau(H-E_T)}\Phi(\mathbf{r},0),
\end{equation}
acts as a projector onto the lowest state compatible with the nodal pattern of the initial condition. The reference energy $E_T$ shifts the origin of the spectrum in the propagator and serves as a slow control parameter that keeps the overall normalization of $\Phi$ under numerical control. For sufficiently large $\tau$, all higher energy components are exponentially damped and $\Phi$ approaches the fixed node ground state in that sector, as in other GFMC style calculations of nonrelativistic QCD bound states~\cite{assi2023baryons,assi2024tetraquarkssufficientlyunequalmassheavy}.

To make this evolution tractable, importance sampling with a guiding function $\Psi_T$ is introduced and the mixed distribution is propagated as
\begin{equation}
f(\mathbf{r},\tau)=\Psi_T(\mathbf{r})\,\Phi(\mathbf{r},\tau),
\end{equation}
which satisfies a drift diffusion equation with a branching term. In its stochastic representation, one short time step is written as a drift diffusion move followed by a weight update. The move is
\begin{equation}
\mathbf{r}'=\mathbf{r}+\mathbf{v}_D(\mathbf{r})\,\Delta\tau
+\sqrt{2D\,\Delta\tau}\,\boldsymbol{\eta},
\end{equation}
where $\boldsymbol{\eta}$ is a vector of independent standard normal variables and
\begin{equation}
D=\frac{1}{2\mu}
\end{equation}
is the diffusion constant fixed by the kinetic term in the Hamiltonian.

The drift velocity steers walkers toward regions where the guiding function is large,
\begin{equation}
\mathbf{v}_D(\mathbf{r})=2\nabla \ln |\Psi_T(\mathbf{r})|,
\end{equation}
which reduces the variance compared to unguided diffusion.

After the move, the walker acquires a weight factor,
\begin{equation}
w\approx \exp\!\Big[-\tfrac{\Delta\tau}{2}\big(
E_L(\mathbf{r}')+E_L(\mathbf{r})-2E_T\big)\Big],
\end{equation}
where the local energy is defined as
\begin{equation}
E_L(\mathbf{r})=\frac{H\Psi_T(\mathbf{r})}{\Psi_T(\mathbf{r})}.
\end{equation}

The weight $w$ reflects how the local energy along the step compares with the reference $E_T$, so walkers that spend more time in regions where $E_L < E_T$ accumulate large weights, while those that mostly sample $E_L > E_T$ are strongly suppressed. This drift diffusion branching structure mirrors the algorithms used in Refs.~\cite{assi2023baryons,assi2024tetraquarkssufficientlyunequalmassheavy} for nonrelativistic QCD Hamiltonians.

In the actual calculation, spherical symmetry is exploited so that only the radial coordinate is propagated. The angular dependence is carried entirely by spherical harmonics and absorbed into normalization factors. The move above therefore reduces to an update of the scalar radius $r$, driven by the radial component of the drift velocity and a one dimensional Gaussian increment with variance $2D\,\Delta\tau$. Reflecting boundary conditions at $r=0$ and $r=r_{\max}$ are imposed by mirroring any proposed step that leaves the interval $[0,r_{\max}]$ back into the domain. This procedure enforces regularity at the origin and vanishing amplitude at large radius without introducing artefacts.

Population control is handled by combining the continuous weights $w$ with occasional stochastic reconfiguration (branching). After several drift diffusion steps, walkers with very small weights are removed and walkers with very large weights are split into several copies with proportionally reduced weights, so that the effective ensemble size remains close to a target population $M_0$ and statistical estimates are not dominated by a few outliers. During an initial equilibration period the reference energy $E_T$ is adjusted slowly, based on the observed growth or decay of the total weight or walker count, to prevent runaway behaviour. Once the population and the mixed estimator have settled onto a plateau, $E_T$ is frozen and the subsequent evolution is used purely for projection and measurement, with no further feedback that could bias the energy, following the same philosophy as in other fixed node projector studies~\cite{assi2023baryons,assi2024tetraquarkssufficientlyunequalmassheavy}.

Energies are estimated with the mixed estimator
\begin{equation}
E_{mix}(\Delta\tau)=\frac{\langle \Psi_T|H|\Phi\rangle}{\langle \Psi_T|\Phi\rangle}
=\Big\langle E_L(\mathbf{r})\Big\rangle_{f(\mathbf{r})},
\end{equation}
which converges to the fixed node ground state energy in the long time limit. For operators that do not commute with $H$, the usual extrapolated estimator $2\langle O\rangle_{\mathrm{mix}}-\langle O\rangle_{\mathrm{var}}$ is employed, and in representative cases agreement has been checked with descendant weighting (forward walking) constructions, as is standard in projector Monte Carlo calculations~\cite{assi2023baryons}. Short time bias is removed by running a small ladder of time steps $\{\Delta\tau_i\}$ in a regime where the dynamics is stable and fitting
\begin{equation}
E_{mix}(\Delta\tau_i)=E_0+a\,\Delta\tau_i+b\,(\Delta\tau_i)^2,
\end{equation}
with $E_0$ taken as the $\Delta\tau\to 0$ limit. For each $\Delta\tau_i$ the total projection time $\tau_{tot}=N_{steps}\Delta\tau_i$ is chosen large enough that transients from the initial condition have clearly decayed, and the same $\tau_{tot}$ is maintained approximately constant as $\Delta\tau$ is varied by adjusting $N_{steps}$. Statistical uncertainties on $E_0$ are obtained from blocked time series and cross-checked by omitting the largest $\Delta\tau$ point or repeating runs at increased walker populations, in close analogy to the stability checks performed in Refs.~\cite{assi2023baryons,assi2024tetraquarkssufficientlyunequalmassheavy}.

Calibration of the potential parameters then proceeds by comparing the projected energies to spin-averaged experimental mass of ground state, as discussed in Sec.~\ref{sec:calibration}. Both absolute masses and splittings relative to the $1S$ state are considered. In either case, parameter scans are performed on a grid in $(\sigma,\kappa)$, the constant $V_0$ is fixed by a simple anchoring convention, and only those grid points for which VMC and GFMC agree within uncertainties and the $\Delta\tau\to 0$ extrapolation is stable are retained for further analysis. Masses are finally assembled via Eq.~\eqref{eq:mass}, while quoted splittings are taken directly from the projected energies and are independent of $V_0$.

\section{Reference Inputs, Calibration Strategy and GFMC Stability}
\label{sec:calibration}

The calibration of the Cornell-type potential is carried out using the representative literature values for the spin averaged tower of the $B_c$ spectrum. The analysis includes the ground state and a ladder of low-lying radial and orbital excitations, from the $1S$ ground state up to the first few excited $S$-, $P$-, and $D$-waves. At present, the $1S$ centroid is the only $B_c$ spin-averaged level regarded as a robust experimental input for calibration, and it therefore serves as the anchor for the Cornell parameters. For higher levels, there is no attempt to directly fit the sparse experimental information. Instead, the predicted masses are compared with representative values collected from the literature~\cite{cdf1998observation,abazov2008measurement,lhcb2015measurement,aaij2018measurement,aaij2023study,allison2005mass,beringer2012review,patrignani2016review}.

On the theory side the dynamics are described by the Cornell-type potential introduced in Eq.~\eqref{eq:veff}, with string tension $\sigma$, short distance Coulomb strength $\kappa$, and an additive constant $V_0$ that fixes the absolute energy origin. The heavy quark masses are fixed to $m_c = 1.2730\,GeV$ and $m_b = 4.183\,GeV$ \cite{ParticleDataGroup:2024cfk}. With these inputs held fixed, the calibration constrains only the pair $(\sigma,\kappa)$, while $V_0$ is determined by anchoring the $1S$ state as described below. This strategy mirrors many phenomenological studies of heavy quarkonia and $B_c$ mesons, in which potential parameters are tuned to reproduce selected reference
levels~\cite{eichten1994mesons,godfrey1985mesons,ebert2003properties,godfrey2004spectroscopy,ebert2011spectroscopy,soni2018qq,devlani2014masses,monteiro2017cb,li2023higher,li2023spectroscopic}.

For each choice of $(\sigma,\kappa)$ the spectrum is obtained in two stages. First, a VMC calculation produces an optimized trial wave function $\Psi_T$ for each state in the $B_c$ tower. The VMC parameters are tuned to minimize the expectation value of the Hamiltonian. This step is repeated at every point of the $(\sigma,\kappa)$ scan, ensuring that the trial state remains adapted to the potential. The resulting $\Psi_T$ is then used as the importance sampling distribution and as the nodal template for the GFMC stage, in close analogy to the role of trial states in recent nonrelativistic QCD applications of VMC+GFMC~\cite{assi2023baryons,assi2024tetraquarkssufficientlyunequalmassheavy}.

In the second stage a fixed node GFMC evolution projects the VMC trial state in imaginary time and removes the residual variational bias. For each channel $(n,\ell)$ in the $B_c$ tower, a projected eigenvalue $E_{n\ell}(\sigma,\kappa,V_0)$ is obtained. The spin-averaged mass in Eq.~\eqref{eq:mass} then depends on the parameters of the potential as
\begin{equation}
M^{\mathrm{th}}_{n\ell}(\sigma,\kappa,V_0)
= m_b + m_c + E_{n\ell}(\sigma,\kappa,V_0).
\label{eq:mass_theory}
\end{equation}
It is often convenient to work with level spacings relative to the ground state centroid,
\begin{equation}
\Delta_{n\ell} = M_{n\ell} - M_{1S}, \qquad
\Delta^{\mathrm{th}}_{n\ell} = M^{\mathrm{th}}_{n\ell}-M^{\mathrm{th}}_{1S},
\qquad
\Delta^{\mathrm{exp}}_{n\ell} = M^{\mathrm{exp}}_{n\ell}-M^{\mathrm{exp}}_{1S},
\label{eq:delta_def}
\end{equation}
which are independent of $V_0$ and probe only the shape of the potential. These splittings will be used later when discussing Fig.~\ref{fig:splitting_bars}, but they are not needed explicitly in
the numerical score. Similar use of splittings to minimise sensitivity to additive constants is standard in potential model analyses of heavy quark spectra~\cite{eichten1994mesons,soni2018qq,devlani2014masses,li2023higher,li2023spectroscopic}.

The constant $V_0$ itself is not treated as an independent fit parameter. Instead, for each pair $(\sigma,\kappa)$ it is fixed by requiring that the theoretical $1S$ centroid reproduces the experimental one,
\begin{equation}
M^{\mathrm{th}}_{1S}(\sigma,\kappa,V_0) = M^{\mathrm{exp}}_{1S}.
\end{equation}
To make this condition explicit, the $1S$ level is first computed with $V_0=0$ and denoted by $E_{1S}(\sigma,\kappa,V_0{=}0)$. Using Eq.~\eqref{eq:mass_theory}, the required offset is then
\begin{equation}
V_0(\sigma,\kappa) = M^{\mathrm{exp}}_{1S} - m_b - m_c - E_{1S}(\sigma,\kappa,V_0{=}0).
\label{eq:v0_def}
\end{equation}
This equation determines a unique $V_0$ for each point on the $(\sigma,\kappa)$ grid. In other words, the absolute energy scale of the Cornell potential is calibrated so that the $1S$ mass, which is experimentally known with high
precision~\cite{cdf1998observation,abazov2008measurement,lhcb2015measurement,aaij2018measurement,aaij2023study,beringer2012review,patrignani2016review}, is reproduced exactly; all excited levels are then predictions of this
anchored Hamiltonian. A similar anchoring of $V_0$ to a reference state is used in many Cornell-type fits to charmonium, bottomonium, and $B_c$ spectra~\cite{eichten1994mesons,soni2018qq,devlani2014masses,ebert2011spectroscopy,monteiro2017cb,pathak2022parameterisation,solomko2023cornell}.

Once $V_0(\sigma,\kappa)$ has been determined from Eq.~\eqref{eq:v0_def}, the anchored spectrum $M^{\mathrm{th}}_{n\ell}(\sigma,\kappa,V_0(\sigma,\kappa))$ is compared with the experimental centroids. The agreement between theory and experiment is quantified by a single root mean square error constructed from the absolute masses,
\begin{equation}
\text{RMSE}(\sigma,\kappa) =\left[\frac{1}{N}\sum_{(n,\ell)\in\mathcal{F}}
\big(M^{\mathrm{th}}_{n\ell}-M^{\mathrm{exp}}_{n\ell}\big)^2 \right]^{1/2},
\label{eq:rmse_abs}
\end{equation}
where $\mathcal{F}$ denotes the subset of the $B_c$ tower used in the fit and $N$ is the number of these fit channels. The $1S$ level belongs to $\mathcal{F}$ but contributes trivially, because it is exactly matched by construction through Eq.~\eqref{eq:v0_def}. This kind of parameter-space survey using a global RMSE over selected
levels is closely related to the explorations of Cornell parameter space in
Refs.~\cite{solomko2023cornell,pathak2022parameterisation,sreelakshmi2022mass,ahmad2025charmonium}.

\begin{figure}[H]
\centering
\includegraphics[width=0.82\linewidth]{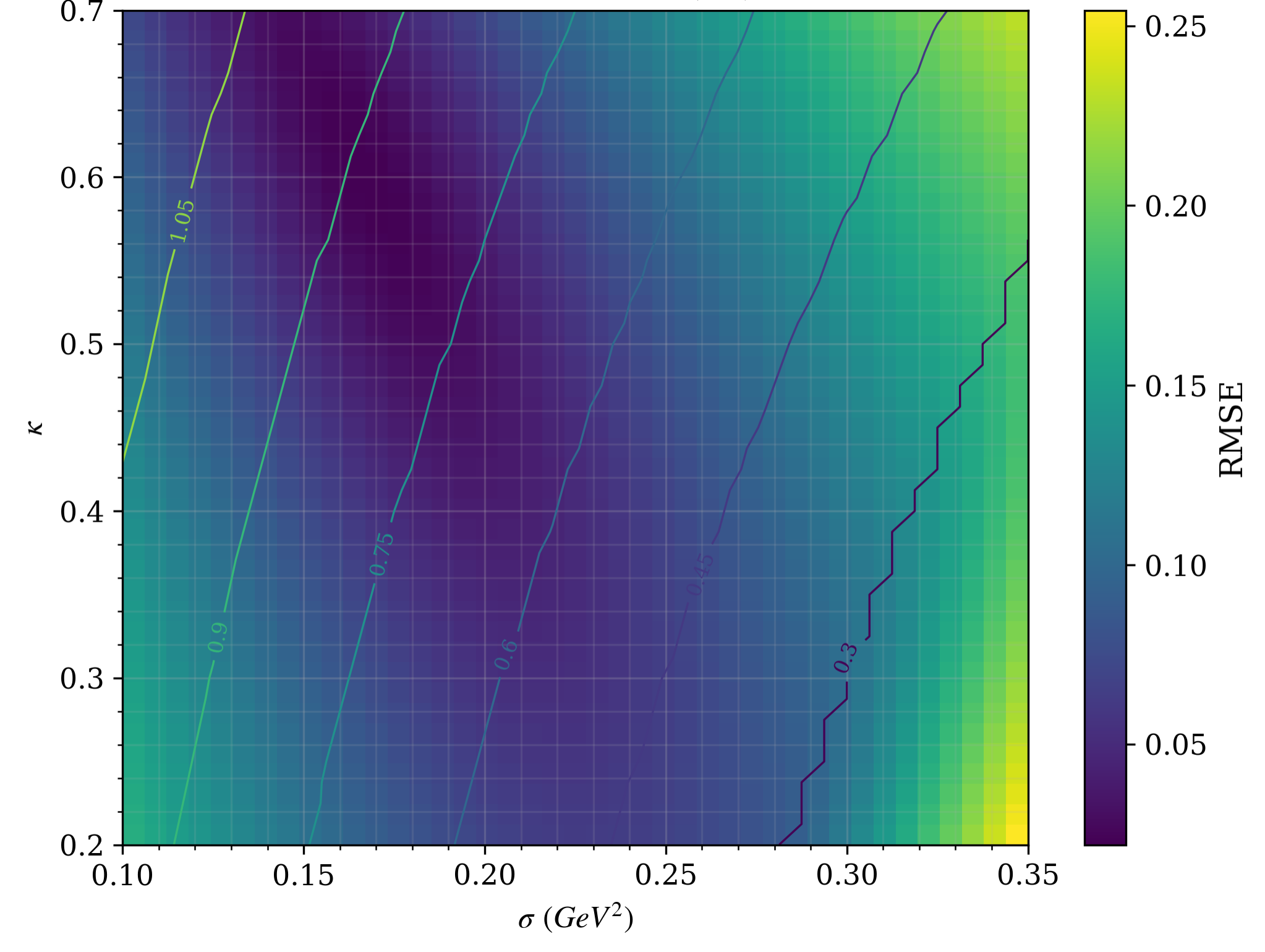}
\caption{RMSE surface in the $(\sigma,\kappa)$ plane with
$V_0(\sigma,\kappa)$ contours. The dark band indicates the low-RMSE
valley, including the best point of Eq.~\eqref{eq:best_point}.}
\label{fig:rmse_heatmap}
\end{figure}

The resulting RMSE surface, together with the corresponding values of $V_0(\sigma,\kappa)$, is shown in Fig.~\ref{fig:rmse_heatmap}. The horizontal axis gives the string tension $\sigma$ and the vertical axis the Coulomb strength $\kappa$. The color scale encodes the value of the RMSE, while the overlaid contour lines display $V_0(\sigma,\kappa)$ in $\mathrm{GeV}$. A broad diagonal valley of low-RMSE is clearly visible. Moving along this valley, a decrease in $\sigma$ can be compensated by an increase in
$\kappa$, and vice versa. This reflects the competition between the linear confining term and the short distance Coulomb attraction. At the same time the $V_0$ contours show that the offset varies smoothly along the
valley and remains in a natural range. All working values of $\sigma$, $\kappa$, and $V_0$ used in the rest of the paper are taken from this figure. In particular, a representative best point in the valley is
\begin{equation}
\sigma = 0.1625~\mathrm{GeV}^2,\qquad
\kappa = 0.6125,\qquad
V_0 = 0.901875~\mathrm{GeV},
\label{eq:best_point}
\end{equation}
which is used for the detailed mass comparisons in Sec.~\ref{sec:results}. When projected onto the same
$(\sigma,\kappa)$ plane, parameter sets extracted for charmonium and bottomonium in
Refs.~\cite{eichten1994mesons,soni2018qq,li2023higher,li2023spectroscopic,solomko2023cornell,pathak2022parameterisation,sreelakshmi2022mass,ahmad2025charmonium,mutuk2019cornell,akan2025predicting}
lie naturally inside or close to this diagonal valley, indicating that the best-fit region is consistent with broader heavy quarkonium phenomenology.

Because the calibration relies directly on GFMC energies, it is essential to understand how the Monte Carlo control parameters and the radial discretisation affect the results. The most important numerical inputs are the imaginary time step $\Delta\tau$, the number of time steps $N_{steps}$ (and hence the total projection time), the walker population, and the radial grid $(r_{\max},N_r)$. For a given channel and choice of $(\sigma,\kappa,V_0)$, the GFMC evolution uses a short time Green’s function with leading discretization errors of order $\Delta\tau^2$, as in other fixed node projector studies of nonrelativistic QCD systems~\cite{assi2023baryons,assi2024tetraquarkssufficientlyunequalmassheavy}.

\begin{figure}[H]
\centering
\subfloat[]{
  \includegraphics[width=0.48\linewidth]{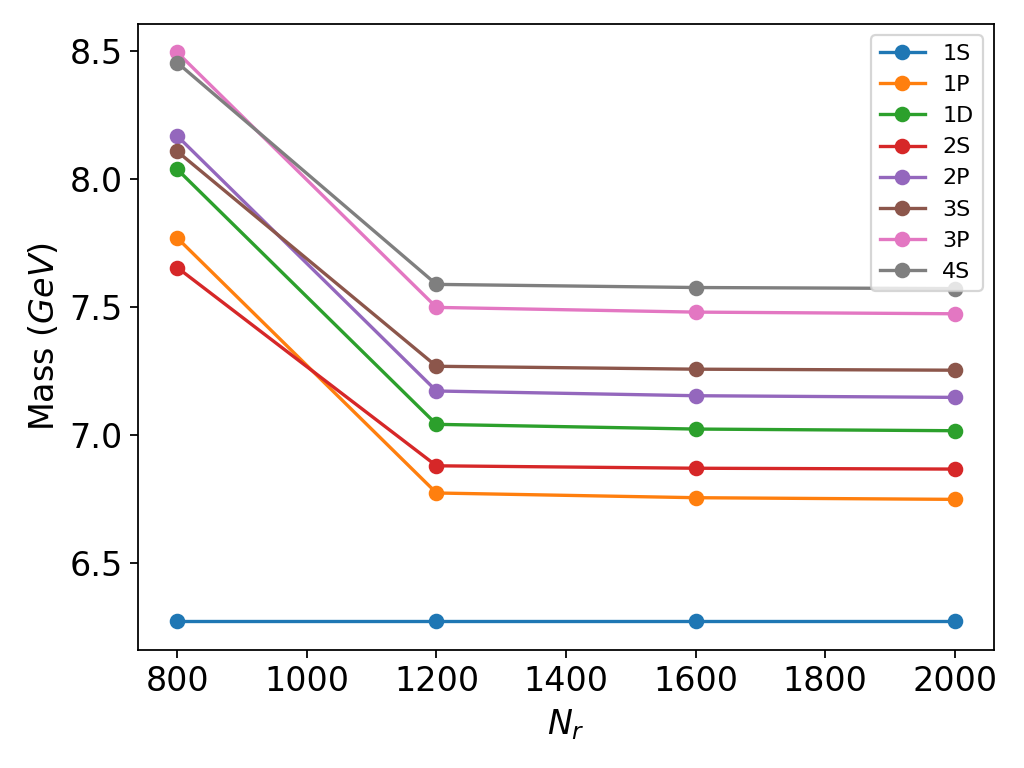}
  \label{fig:nr_scan}
}
\hfill
\subfloat[]{
  \includegraphics[width=0.48\linewidth]{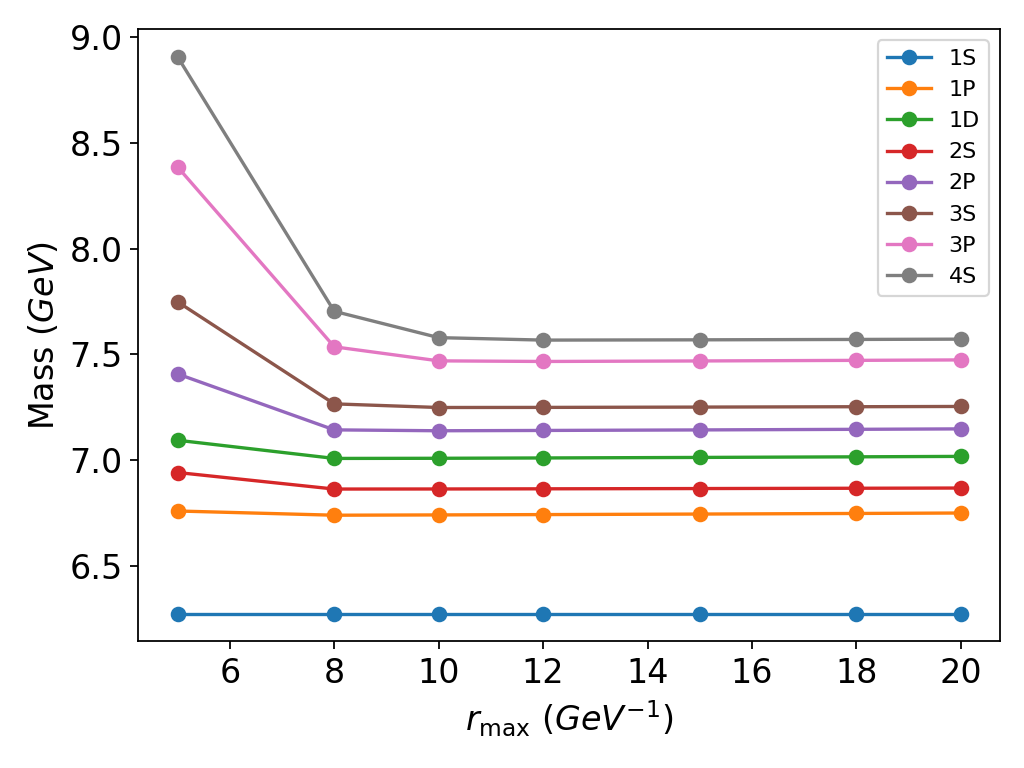}
  \label{fig:rmax_scan}
}
\caption{Spin-averaged $B_c$ masses at the best valley parameters as functions of (a) the number of radial grid points $N_r$ at fixed $r_{\max}=15~\mathrm{GeV}^{-1}$ and (b) the radial cutoff $r_{\max}$ at fixed $N_r=10^3$.}
\label{fig:nr_rmax_scans}
\end{figure}

To control radial discretisation and finite volume effects, the number of radial grid points $N_r$ is scanned at fixed $r_{\max}=15~\mathrm{GeV}^{-1}$ and, in a separate sweep, the radial cutoff $r_{\max}$ is scanned at fixed $N_r=10^3$. The resulting spectra are shown in Fig.~\ref{fig:nr_rmax_scans}. In both panels the spin-averaged masses for all members of the tower rapidly approach a common plateau: variations become negligible once $N_r$ and $r_{\max}$ exceed these reference values, which are therefore adopted in all production runs.

The mixed estimator of the energy is then measured at several time steps $\Delta\tau_i$ while $N_{steps}$ and the other control parameters are held fixed. The total projection time is
\begin{equation}
\tau_{tot} = N_{steps}\,\Delta\tau.
\label{eq:tau_tot_def}
\end{equation}
If $\Delta\tau$ is chosen too small at fixed $N_{steps}$, then $\tau_{tot}$ is not long enough for the projection to reach the asymptotic plateau and the energy estimates remain biased towards the higher VMC values. In this regime the results are systematically high, even though the formal $\Delta\tau^2$ error of
the short time propagator is small. On the other hand, if $\Delta\tau$ is too large, discretization errors from the short-time approximation dominate and spoil the continuum limit.

\begin{figure}[H]
\centering
\includegraphics[width=0.94\linewidth]{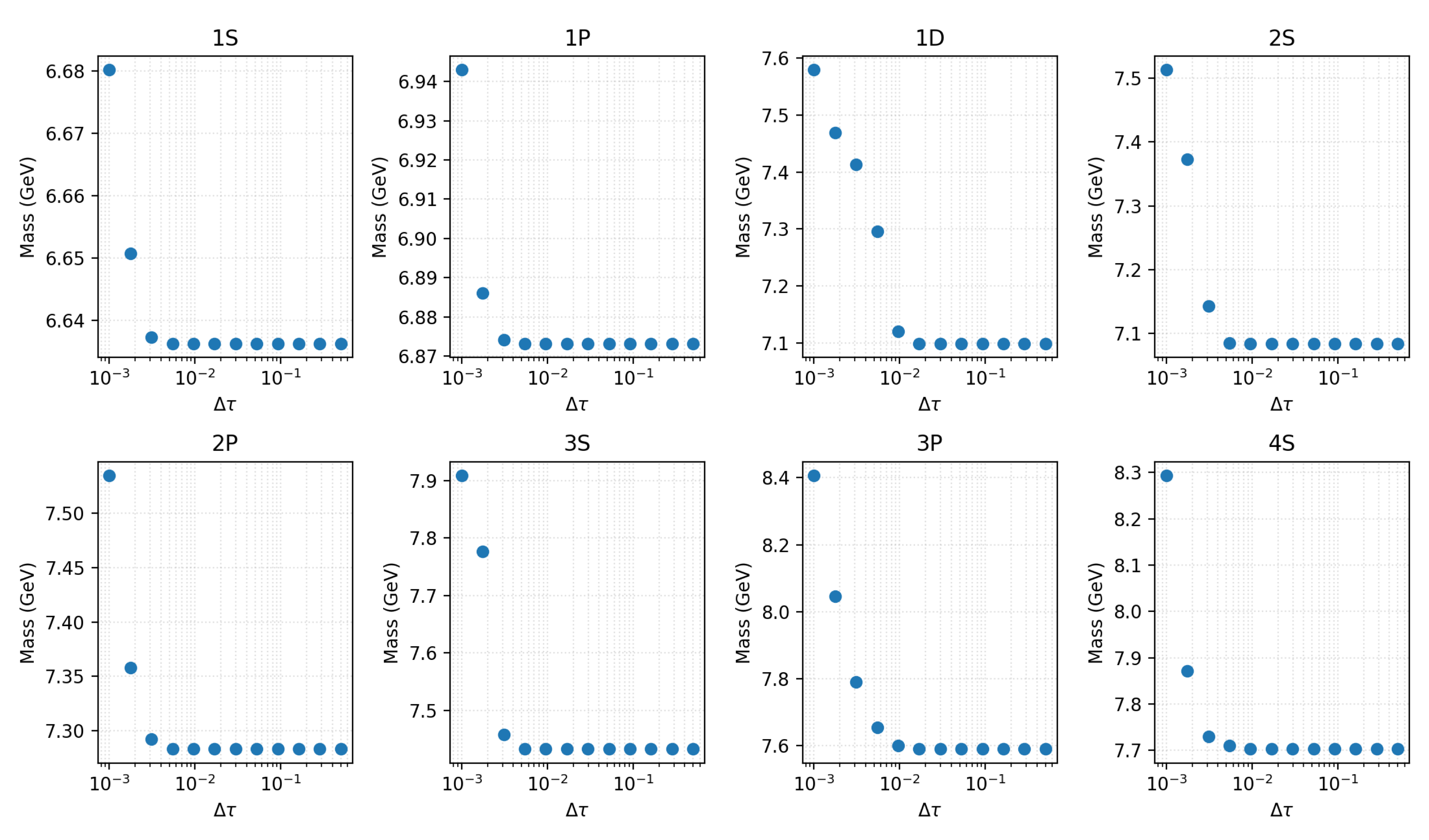}
\caption{Spin-averaged $B_c$ masses versus time step $\Delta\tau$ at fixed $N_{steps}=10^4$.}
\label{fig:dtau}
\end{figure}

In practice, an intermediate window of $\Delta\tau$ values is sought in which both conditions are satisfied, namely that $\tau_{\text{tot}}$ is long enough for the projection to converge and $\Delta\tau$ is small enough that short time errors remain mild. In this window the mixed estimator $E_{mix}(\Delta\tau)$ is approximately constant. This behaviour is illustrated in Fig.~\ref{fig:dtau}, which shows the spin-averaged masses for the $B_c$ tower as functions of $\Delta\tau$ in a logarithmic sweep at fixed $N_{steps}$. At the smallest steps the projection time is too short and the masses are biased upward; at the largest steps the time step error becomes visible. The central plateau, where all levels in the tower are consistent with a common value within uncertainties, defines the
useful window in $\Delta\tau$. Within this window the time step dependence is fitted to the quadratic form
\begin{equation}
E_{mix}(\Delta\tau) = E_0 + a\,\Delta\tau + b\,\Delta\tau^2,
\label{eq:etau_fit}
\end{equation}
and the extrapolated value $E_0$ is taken as the continuum time estimate of the energy that enters Eq.~\eqref{eq:rmse_abs}. The walker population and the radial grid $(r_{\max},N_r)$ are increased until the changes in $E_0$ are negligible at the quoted precision, so that statistical noise and finite volume effects do not distort the calibration, following the same philosophy as in Refs.~\cite{assi2023baryons,assi2024tetraquarkssufficientlyunequalmassheavy}.

\begin{figure}[H]
\centering
\includegraphics[width=0.94\linewidth]{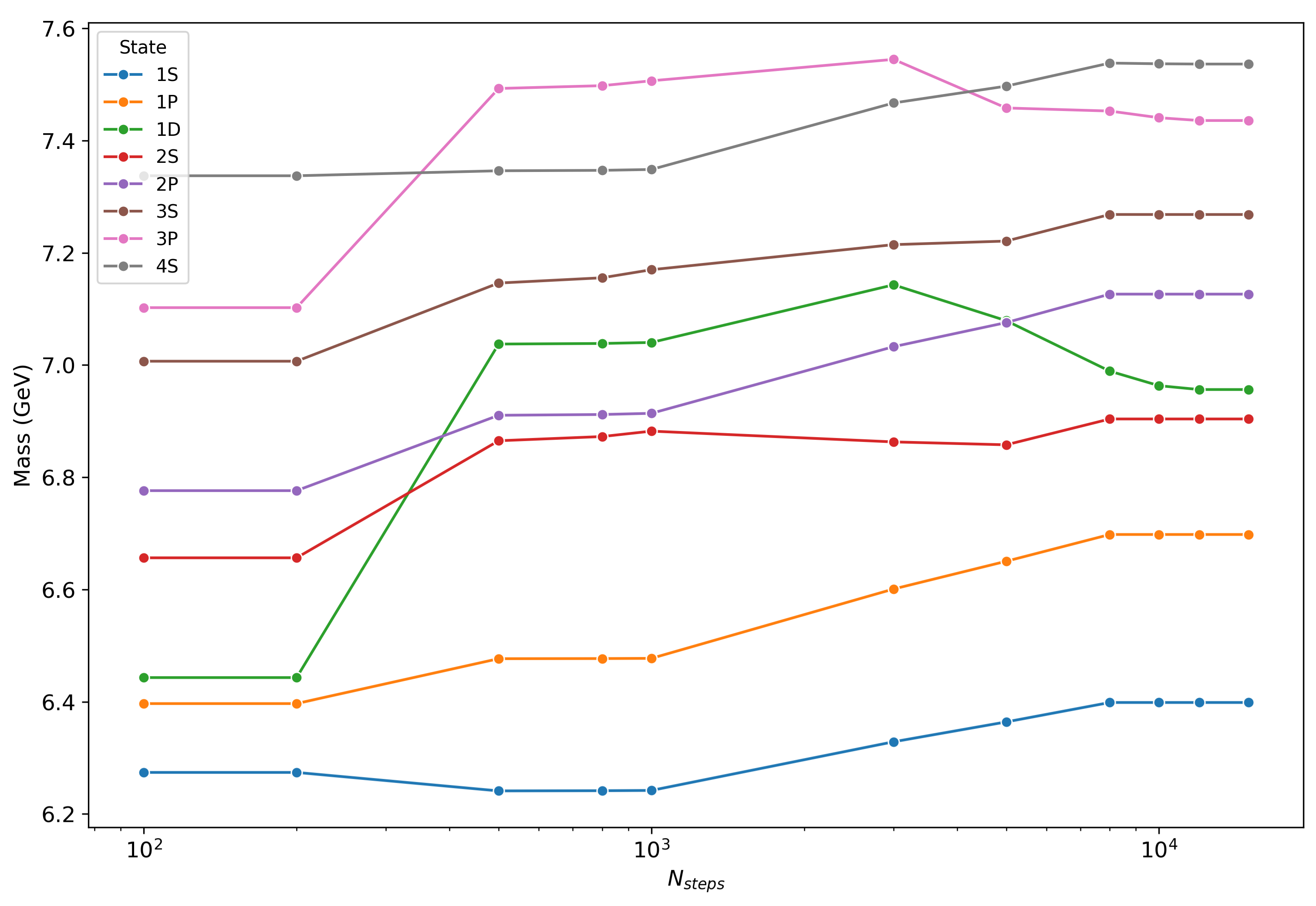}
\caption{Spin-averaged $B_c$ masses versus number of GFMC steps
$N_{steps}$ at fixed $\Delta\tau = 10^{-3}$. All states approach a
common plateau for $N_{steps}\gtrsim 10^{4}$.}
\label{fig:nsteps}
\end{figure}

A complementary test of projection quality is obtained by varying $N_{steps}$ at fixed $\Delta\tau = 10^{-3}$, chosen from the central plateau in Fig.~\ref{fig:dtau}, as shown in Fig.~\ref{fig:nsteps}. Starting from the VMC trial state, the spin-averaged masses for all levels in the $B_c$ tower fall from their variational values and then
evolve as $\tau_{tot} = N_{steps}\Delta\tau$ is increased. At relatively small values of $N_{steps}$, around $10^{2}$, the masses form a smooth pattern from the ground state up through the higher excitations. As $N_{steps}$ grows, each trajectory drifts towards its projected value and the relative position of nearby excited levels is
modified: for instance, between $N_{steps}\sim10^{2}$ and $N_{steps}\sim10^{4}$ the $1D$ mass moves slightly above the $2S$ mass, and the $4S$ mass moves above the $3P$ mass. These changes show that increasing $N_{steps}$ does not simply translate all levels rigidly but probes the imaginary time dynamics of each channel: some states relax faster, others more slowly, and the intermediate values of $N_{steps}$ still contain transient contributions from higher excitations. By $N_{steps}\sim 10^{4}$ the drift of all curves has essentially saturated, and independent fits to individual states yield mutually compatible plateau values within the quoted uncertainties. All production runs used in the calibration are performed in this long time window. The residual differences between the last two $N_{steps}$ values, typically at most a few tens of MeV for any given state, are interpreted as an upper bound on the remaining projection systematic, in line with the convergence criteria adopted in other GFMC studies~\cite{assi2023baryons,assi2024tetraquarkssufficientlyunequalmassheavy}.

Within the low-RMSE valley we further discriminate between nearby points by looking at the residuals for each state in the $B_c$ tower. Points where the deviations from experiment drift systematically with excitation (for example, progressively overshooting the higher radial states) are disfavoured even if their global RMSE is small. Instead, parameter sets are selected such that all fitted levels remain close to the data without an obvious tilt across the tower. A point near the middle of the valley, given explicitly in Eq.~\eqref{eq:best_point}, satisfies these criteria and is used as the baseline. This kind of level by level scrutiny of residuals is similar in spirit to the detailed comparison of potential models and lattice spectra in~\cite{eichten1994mesons,ebert2003properties,ebert2011spectroscopy,li2023higher,li2023spectroscopic,Chen_2020,chaturvedi2022b}.

\begin{figure}[H]
\centering
\includegraphics[width=0.92\linewidth]{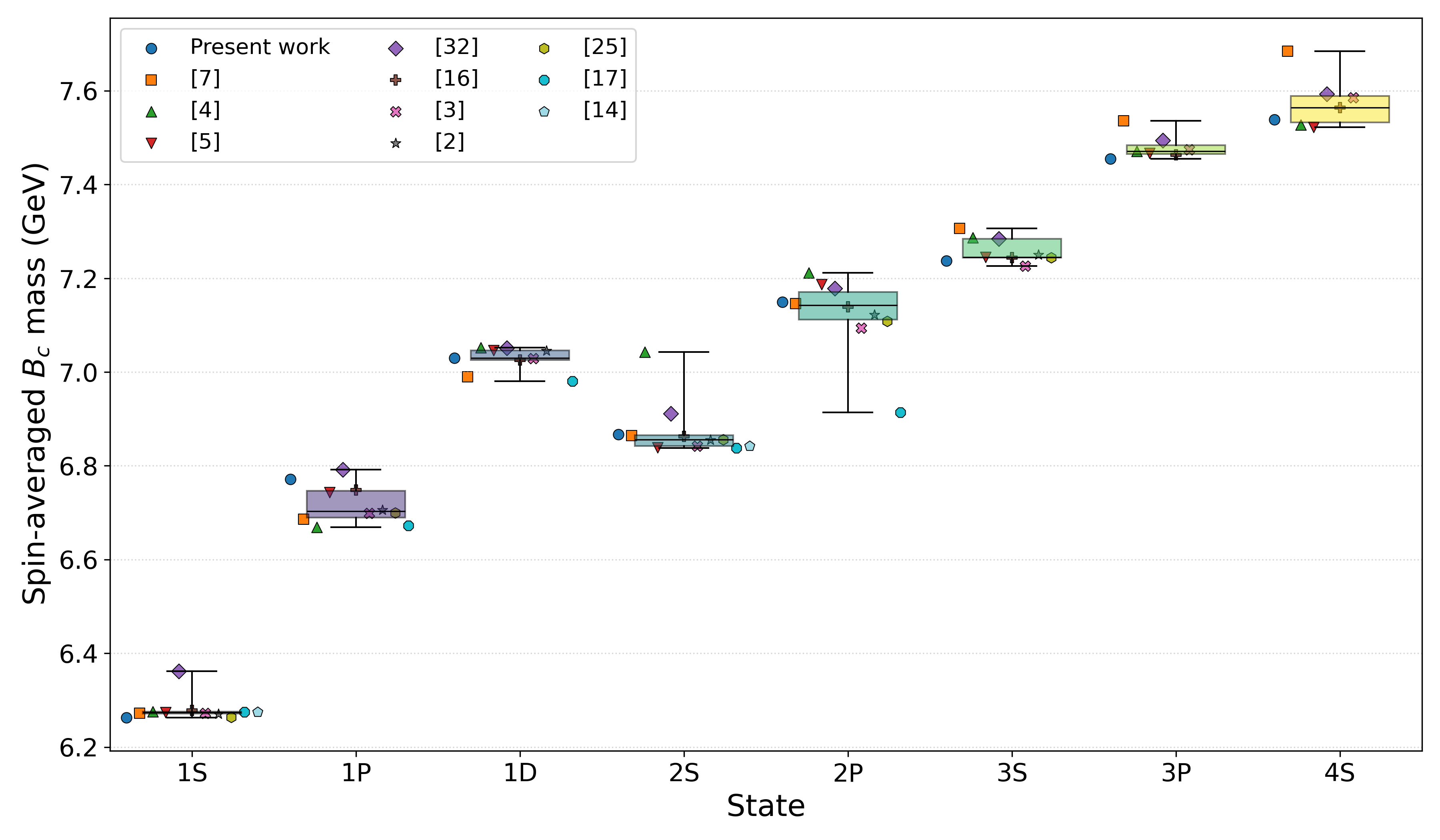}
\caption{Spin-averaged $B_c$ masses for each state in the tower. For every level, the box and whisker band summarises the spread of representative literature values, together with available experimental information where present.}
\label{fig:mass_bars}
\end{figure}

\begin{figure}[H]
\centering
\includegraphics[width=0.92\linewidth]{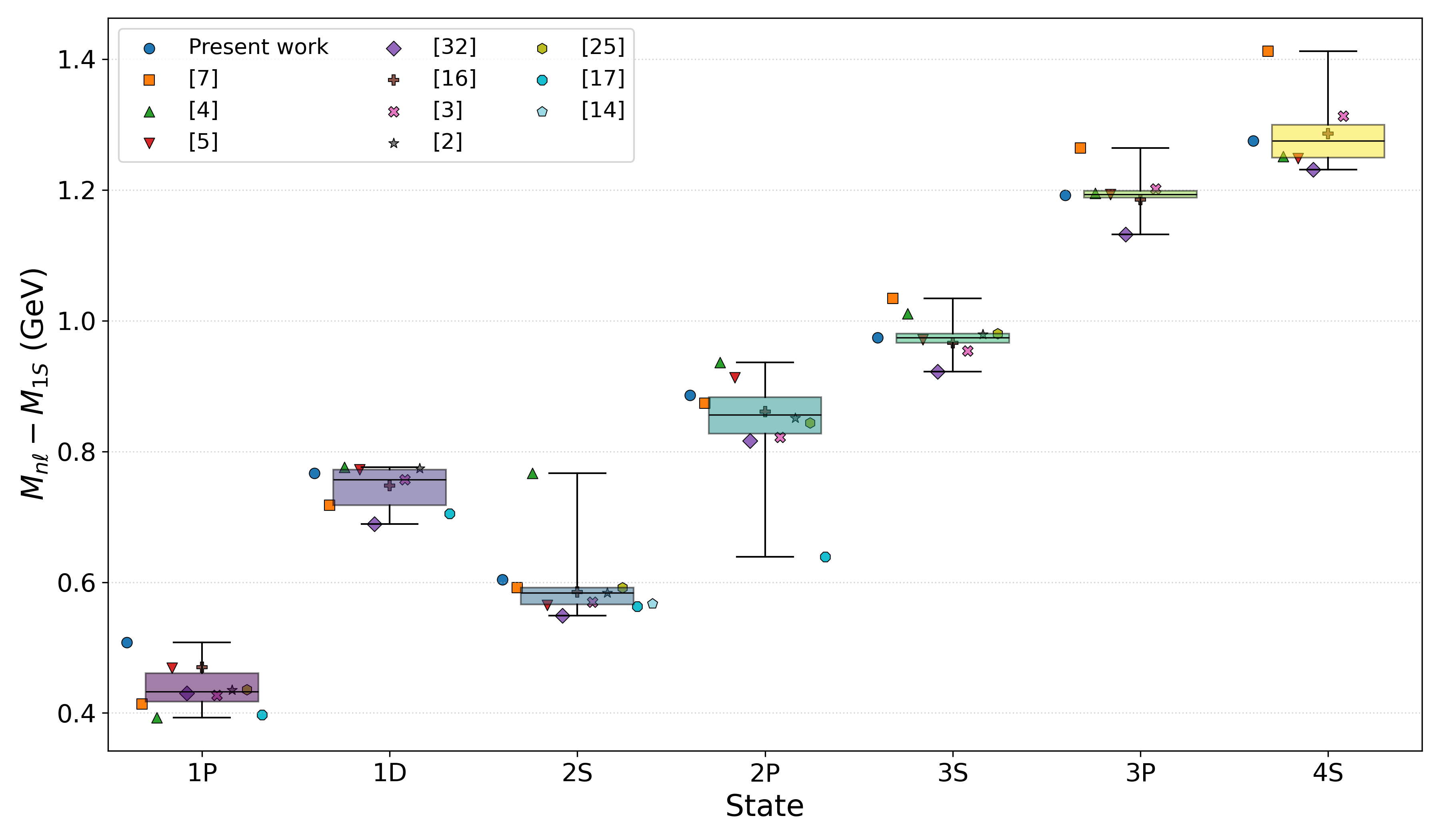}
\caption{Same as Fig.~\ref{fig:mass_bars}, shown as splittings relative to the $1S$ centroid using Eq.~\eqref{eq:delta_def}. The boxes and whiskers again represent the spread of literature values, not fit uncertainties for a single calculation.}
\label{fig:splitting_bars}
\end{figure}

The quality of the calibration at this representative valley point is summarised in Figs.~\ref{fig:mass_bars} and~\ref{fig:splitting_bars}. In these plots each group of box and whisker bands corresponds to one level in the $B_c$ tower. For a given state, the coloured box shows the interquartile range (25--75\,\%) of all available theoretical and experimental determinations collected from the literature, the horizontal line inside the box is the median, and the whiskers extend to the minimum and maximum values. The individual determinations are overlaid as separate markers and are identified in the legend as follows: the GFMC results (``Present work''), Cornell-type potential models of \cite{soni2018qq,devlani2014masses,eichten1994mesons,monteiro2017cb}; the neural network Cornell+quadratic model dedicated to $b\bar c$ spectroscopy~\cite{akan2025predicting} and the purely Cornell based
neural network solver~\cite{mutuk2019cornell}; eigenvalues obtained with the asymptotic iteration method
(AIM)~\cite{ciftci2003asymptotic,kumar2013asymptotic}; relativised quark model spectra of~\cite{ebert2011spectroscopy,godfrey2004spectroscopy}; and the experimental centroid~\cite{patrignani2016review}. The bars in these figures should therefore not be interpreted as statistical uncertainties on a single calculation; instead, they visualise the spread of published values for each state. The underlying numerical values for these determinations are collected in Table~\ref{tab:other_works_comparison} in Sec.~\ref{sec:results}. Similar comparative surveys of multiple approaches to the $B_c$ spectrum can be found in, for example, Refs.~\cite{li2023higher,li2023spectroscopic,chaturvedi2022b,Chen_2020}.

Several qualitative features are apparent from these figures. First, the $1S$ boxes are very narrow, reflecting the fact that all models are constrained, either explicitly or implicitly, to reproduce the measured ground state mass. The situation becomes more discriminating for excited levels. For the $1P$ and $1D$ states all approaches lie
within a band of order $20$--$30~\mathrm{MeV}$, with the GFMC result close to the median of the distribution. The $2S$ and $2P$ levels exhibit the largest spread: the Cornell+quadratic neural-network model~\cite{akan2025predicting} tends to predict somewhat heavier radials, while the relativistic quark model calculations~\cite{ebert2011spectroscopy,godfrey2004spectroscopy} yield a slightly more compressed spectrum. The purely Cornell based neural network calculation of~\cite{mutuk2019cornell}, the AIM results~\cite{ciftci2003asymptotic,kumar2013asymptotic}, and the potential model studies~\cite{soni2018qq,devlani2014masses,eichten1994mesons,monteiro2017cb} track one another closely and sit near the centre of the interquartile bands. The fixed node GFMC points follow the median trend in both masses and splittings and remain within the interquartile band of the literature for all states in the tower. This pattern is consistent with the broader comparisons presented in~\cite{li2023higher,li2023spectroscopic,chaturvedi2022b,Chen_2020}.

From the perspective of methodology, the comparison underscores that very different solution strategies for the Schrödinger (or quasipotential) equation — traditional potential models, semi-analytic AIM techniques, neural network solvers, and the present VMC+GFMC projection — all converge to a broadly consistent description of the
$B_c$ spectrum. The remaining spread, typically a few tens of MeV in the lower part of the tower and somewhat larger for higher excitations, provides a realistic measure of the current theoretical uncertainty associated with model choices and numerical implementations across these
approaches~\cite{ebert2011spectroscopy,soni2018qq,mutuk2019cornell,akan2025predicting,li2023higher,li2023spectroscopic,chaturvedi2022b,Chen_2020}.

Finally, it is instructive to compare the calibrated $B_c$ parameters with canonical Cornell fits for other heavy quarkonia. Typical charmonium and bottomonium analyses cluster near $\sigma\simeq 0.18~\mathrm{GeV}^2$ and
$\kappa\simeq 0.50$-$0.55$ (with a model dependent $V_0$), as in~\cite{eichten1994mesons,godfrey1985mesons,solomko2023cornell,pathak2022parameterisation,sreelakshmi2022mass,ahmad2025charmonium,li2023higher,li2023spectroscopic,mutuk2019cornell}. The best-fit point for $B_c$, Eq.~\eqref{eq:best_point}, lies in a nearby region of the plane and corresponds, via Eq.~\eqref{eq:v0_def}, to an offset $V_0\simeq 0.90~\mathrm{GeV}$. This pattern is consistent with the mixed flavour reduced mass of the $B_c$ system: a slightly stronger short distance interaction helps to place the $S$ waves correctly, while a somewhat softer string tension prevents the radial and orbital ladders in the $B_c$ tower from spreading too far, in line with the qualitative behaviour found in other Cornell-type parameterisations~\cite{solomko2023cornell,pathak2022parameterisation,sreelakshmi2022mass,ahmad2025charmonium}.

\section{Results and Discussion}
\label{sec:results}

At the representative best point in Eq.~\eqref{eq:best_point}, the Cornell GFMC setup yields the spin-averaged $B_c$ spectrum summarised in Table~\ref{tab:other_works_comparison}. The predicted masses track the experimental centroids closely across the full tower; the differences are at the level of a few tens of MeV, which is small on heavy quark scales and well within the accuracy expected of a spin-independent Cornell model. In addition to the present work and
the experimental centroid, the table includes a wide range of potential model calculations for the $B_c$ system, namely Refs.~\cite{soni2018qq,akan2025predicting,mutuk2019cornell,
ciftci2003asymptotic,devlani2014masses,ebert2011spectroscopy,
godfrey2004spectroscopy,eichten1994mesons,monteiro2017cb}, the same set that underlies Figs.~\ref{fig:mass_bars} and~\ref{fig:splitting_bars}, and is representative of modern $B_c$ spectroscopy studies based on nonrelativistic potentials and related approaches~\cite{ebert2003properties,li2023higher,li2023spectroscopic,chaturvedi2022b,Chen_2020}.

\begin{table}[H]
\centering
\caption{Spin-averaged $B_c$ masses (in $GeV$) from the present work compared with other model calculations and available experimental informations.}
\label{tab:other_works_comparison}
\begin{tabular}{lcccccccccccc}
\hline\hline
State 
& Present
& \cite{soni2018qq} 
& \cite{akan2025predicting} 
& \cite{mutuk2019cornell} 
& \cite{ciftci2003asymptotic} 
& \cite{devlani2014masses} 
& \cite{ebert2011spectroscopy} 
& \cite{godfrey2004spectroscopy} 
& \cite{eichten1994mesons} 
& \cite{monteiro2017cb} 
& \cite{patrignani2016review} \\
& work &&&&&&&&&&\\
\hline
$1S$ & 6.263 & 6.272 & 6.276 & 6.274 & 6.362 & 6.278 & 6.272 & 6.271 & 6.264 & 6.275 & 6.275 \\
$1P$ & 6.771 & 6.686 & 6.669 & 6.743 & 6.792 & 6.748 & 6.699 & 6.706 & 6.700 & 6.672 & --    \\
$1D$ & 7.030 & 6.990 & 7.052 & 7.046 & 7.051 & 7.026 & 7.029 & 7.045 & --    & 6.980 & --    \\
$2S$ & 6.867 & 6.864 & 7.043 & 6.839 & 6.911 & 6.863 & 6.842 & 6.855 & 6.856 & 6.838 & 6.842 \\
$2P$ & 7.149 & 7.146 & 7.212 & 7.187 & 7.178 & 7.139 & 7.094 & 7.122 & 7.108 & 6.914 & --    \\
$3S$ & 7.237 & 7.306 & 7.287 & 7.245 & 7.284 & 7.244 & 7.226 & 7.250 & 7.244 & --    & --    \\
$3P$ & 7.455 & 7.536 & 7.471 & 7.467 & 7.494 & 7.463 & 7.474 & --    & --    & --    & --    \\
$4S$ & 7.538 & 7.684 & 7.527 & 7.522 & 7.593 & 7.564 & 7.585 & --    & --    & --    & --    \\
\hline\hline
\end{tabular}
\end{table}

If all eight levels in Table~\ref{tab:other_works_comparison} are included, the deviation between the GFMC results and the experimental centroids is about $22~MeV$, with an average offset of roughly $-2~MeV$, using the average $B_c$ mass as compiled in~\cite{beringer2012review,patrignani2016review} as the experimental input.Given the simplicity of the Hamiltonian (nonrelativistic and spin-independent), such differences are numerically small and are not interpreted as indicating any tension with the data.The lowest states, $1S$, $1P$, and $1D$, which are most directly involved in constraining the potential, are reproduced well, and the first radial excitations $2S$ and $2P$ show deviations of similar size but opposite sign, so there is no clear systematic drift with excitation. The largest difference occurs for the $2P$ level and remains of the same order. Comparison with the other theoretical curves in
Table~\ref{tab:other_works_comparison} and Figs.~\ref{fig:mass_bars}-\ref{fig:splitting_bars} shows that this level of agreement is typical of modern potential-model descriptions of the $B_c$ system~\cite{ebert2003properties,ebert2011spectroscopy,godfrey2004spectroscopy,soni2018qq,devlani2014masses,li2023higher,li2023spectroscopic,chaturvedi2022b,Chen_2020}.

A key feature of the setup is that the $1S$ ground state is used as a baseline both for the potential and for the Monte Carlo control parameters. On the potential side, the $1S$ centroid anchors $V_0$ through Eq.~\eqref{eq:v0_def} and selects the best point \eqref{eq:best_point} in the $(\sigma,\kappa)$ valley, following the anchoring strategy widely used in Cornell-type fits to heavy quarkonia~\cite{eichten1994mesons,soni2018qq,solomko2023cornell,pathak2022parameterisation,sreelakshmi2022mass,ahmad2025charmonium,mutuk2019cornell}. On the GFMC side, the $1S$ channel provides the most stable and precise signal for locating the $\Delta\tau$ plateau and the saturation region in $N_{steps}$ (Figs.~\ref{fig:dtau} and~\ref{fig:nsteps}), in line with standard practice in projector Monte Carlo treatments of nonrelativistic QCD Hamiltonians~\cite{assi2023baryons,assi2024tetraquarkssufficientlyunequalmassheavy}. The time step and projection time windows are therefore tuned using the $1S$ state and then \emph{kept fixed} for the entire $B_c$ tower. In this sense the ground state defines a common numerical and dynamical baseline: once it is under control, the excited levels become genuine predictions of the same Hamiltonian and the same GFMC setup.

The higher members of the tower, $3S$, $3P$, and $4S$, are not needed to define the low-RMSE valley but provide an out of sample test of this baseline. Their masses remain close to the experimental centroids and fall within the spread of the other potential model results, confirming that the combination of $(\sigma,\kappa,V_0)$ in Eq.~\eqref{eq:best_point} and the GFMC parameters extracted from the $1S$ analysis continues to organise the spectrum consistently up the ladder. In particular, the $3S$-$3P$-$4S$ pattern is reproduced without any additional tuning, consistent with the ordering seen in other analyses of higher $B_c$ excitations~\cite{li2023higher,li2023spectroscopic,chaturvedi2022b}.

To put the calibrated Cornell parameters themselves in context, Table~\ref{tab:cornell_params_comparison} compares the best valley values of $(\sigma,\kappa,V_0)$ with representative choices used in selected heavy quarkonium and $B_c$ studies. The classic Cornell fit of \cite{eichten1994mesons} and the soft-wall holographic analysis of \cite{solomko2023cornell} both favour $\sigma\simeq 0.18~GeV^2$ and $\kappa\simeq 0.5$ with a moderately negative constant, while the charmonium fit of \cite{sreelakshmi2022mass} and the neural network Cornell study of \cite{mutuk2019cornell} shift $\sigma$ slightly upward and $\kappa$ slightly downward or upward, respectively, but remain in the same broad band. The screened Cornell potential used by \cite{li2023higher} prefers a noticeably larger string tension and a more negative constant, whereas the charmonium fit of \cite{ahmad2025charmonium} combines a softer string tension with a comparatively strong Coulomb term, illustrating the range of $(\sigma,\kappa,V_0)$ values that can still accommodate heavy quark data.

\begin{table}[H]
\centering
\caption{Comparison of Cornell-type potential parameters
$V(r) = -\kappa/r + \sigma r + V_0$ used in selected heavy quarkonium
and $B_c$ studies.}
\label{tab:cornell_params_comparison}
\begin{tabular}{lccccccc}
\hline
 & Present work 
 & \cite{eichten1994mesons} 
 & \cite{solomko2023cornell} 
 & \cite{sreelakshmi2022mass} 
 & \cite{mutuk2019cornell} 
 & \cite{li2023higher} 
 & \cite{ahmad2025charmonium} \\
\hline
$\sigma$ [GeV$^2$] 
 & 0.1625 & 0.182 & 0.18 & 0.202 & 0.191 & 0.2312 & 0.10 \\
$\kappa$ 
 & 0.6125 & 0.52 & 0.51 & 0.419 & 0.472 & 0.524 & 0.933 \\
$V_0$ [GeV] 
 & 0.9019 & 0 & $-0.30$ & 0.783 & 0 & $-1.171$ & $-0.23$ \\
\hline
\end{tabular}
\end{table}

When these literature parameter sets are projected onto the RMSE surface in Fig.~\ref{fig:rmse_heatmap}, they fall naturally inside or close to the diagonal low-RMSE valley. This confirms that most phenomenologically successful choices of $(\sigma,\kappa,V_0)$ for heavy quarkonia occupy the same region of parameter space that the $B_c$ calibration identifies as optimal, consistent with the broader parameter space surveys in~\cite{solomko2023cornell,pathak2022parameterisation,sreelakshmi2022mass,ahmad2025charmonium}. The best fit point lies comfortably within this corridor: the string tension is somewhat softer than the canonical $0.18~\text{GeV}^2$, the Coulomb strength is modestly larger than $0.5$, and the positive $V_0$ simply reflects the convention adopted here of fixing the absolute mass scale through Eq.~\eqref{eq:v0_def}.

However, the goal of the calibration is not to reproduce any specific set of potential parameters from Table~\ref{tab:cornell_params_comparison}, but to match the measured spectrum as closely as possible. In particular, the $1S$ centroid is a directly observed quantity and is imposed exactly, while the remaining degrees of freedom in $(\sigma,\kappa,V_0)$ are chosen to minimise the RMSE with respect to the experimental spin-averaged masses in
Table~\ref{tab:other_works_comparison}. Agreement with the literature parameter values, as visualised by their location in the heatmap valley, is therefore a valuable consistency check, but the primary criterion is the closeness of the calculated masses — especially the $1S$ state — to the true experimental data~\cite{beringer2012review,patrignani2016review}. In this sense, the spectrum in Figs.~\ref{fig:mass_bars}-\ref{fig:splitting_bars} carries more direct physical weight than the precise numerical value of any single potential parameter.

Taken together, the RMSE valley of Fig.~\ref{fig:rmse_heatmap}, the stability diagnostics of Figs.~\ref{fig:dtau} and~\ref{fig:nsteps}, the detailed comparison in Table~\ref{tab:other_works_comparison}-Figs.~\ref{fig:mass_bars}-\ref{fig:splitting_bars}, and the parameter overview in Table~\ref{tab:cornell_params_comparison} show that a minimal, spin-independent Cornell Hamiltonian, tuned on the $1S$ state and evolved with VMC+GFMC, provides a robust and numerically controlled baseline description of the spin-averaged $B_c$ spectrum. Because the resulting parameters are compatible with canonical charmonium and bottomonium fits~\cite{eichten1978charmonium,eichten1980charmonium,godfrey1985mesons,eichten1994mesons,godfrey2004spectroscopy,solomko2023cornell,pathak2022parameterisation,sreelakshmi2022mass,ahmad2025charmonium}
and lie in the same low-RMSE corridor singled out by other phenomenological studies~\cite{soni2018qq,mutuk2019cornell,li2023higher,li2023spectroscopic,chaturvedi2022b}, this baseline can be reused directly in more elaborate analyses of heavy quark mesons, including spin-dependent splittings, relativistic
corrections, and alternative static potentials, with the present work serving as the reference point against which such refinements can be quantified. In that sense, the VMC+GFMC Cornell solution for $B_c$ complements other nonrelativistic QCD studies of heavy hadrons and multihadron systems~\cite{assi2023baryons}, helping to tie together the spectroscopy of heavy mesons, baryons, and multiquark candidates within a common potential model benchmark.

\bibliographystyle{unsrt}
\bibliography{refs}

\end{document}